\documentclass[english]{article}
\pdfoutput=1
\usepackage{lmodern}
\usepackage{booktabs}
\usepackage[T1]{fontenc}
\usepackage[latin9]{inputenc}
\usepackage{babel}

\makeatletter

\usepackage{mathrsfs}   
\usepackage{slashed}     
\usepackage{url}
\usepackage{graphicx}
\usepackage{color} 
\usepackage[numbers,sort&compress]{natbib}
\usepackage{float}
\usepackage{multirow}
\usepackage{amsmath}
\usepackage{amssymb}
\usepackage{breqn}
\usepackage{subcaption}
\usepackage{caption}
\usepackage{mciteplus}
\usepackage{authblk}
\usepackage{diagbox}
\usepackage{stackengine}
\usepackage[normalem]{ulem}
\newcommand\xrowht[2][0]{\addstackgap[.5\dimexpr#2\relax]{\vphantom{#1}}}

\usepackage{tikz}

\usetikzlibrary{decorations.markings}
\usetikzlibrary{decorations.pathmorphing}

\tikzset{
    vector/.style={decorate, decoration={snake}, draw},
    fermion/.style={draw=black, postaction={decorate},
        decoration={markings,mark=at position .55 with {\arrow[scale=1.,>=stealth]{>}}}},
    fermionbar/.style={draw=black, postaction={decorate},
        decoration={markings,mark=at position .58 with {\arrow[scale=1.,>=stealth]{<}}}},
    fermionline/.style={draw=black},
    gluon/.style={decorate, draw=black,
        decoration={coil,amplitude=4pt, segment length=5pt}},
    scalar/.style={dashed, draw=black, postaction={decorate},
        decoration={markings,mark=at position .55 with {\arrow[scale=1.,>=stealth]{>}}}},
    scalarbar/.style={dashed, draw=black, postaction={decorate},
        decoration={markings,mark=at position .58 with {\arrow[scale=1.,>=stealth]{<}}}},
    scalarline/.style={dashed,draw=black},
    fmassin/.style={draw=black, postaction={decorate},
    decoration={markings, mark=at position .75 with {\arrow[scale=1.,>=stealth]{>}}, mark=at position .30 with {\arrow[scale=1.,>=stealth]{<}}}},
}

\usepackage[colorlinks=true,linkcolor=redLinks,citecolor=greenLinks,urlcolor=redLinks, pdfborder={0 0 1}]{hyperref}
\definecolor{greenLinks}{rgb}{0, 0.6, 0} 
\definecolor{blueLinks}{rgb}{0, 0, 0.6}
\definecolor{redLinks}{rgb}{0.6, 0, 0}
\definecolor{eprintLinks}{rgb}{0.4, 0.4, 0.4}
\definecolor{journalLinks}{rgb}{0.6, 0, 0}

\def\vev#1{\left\langle #1\right\rangle}

\let\oldFootnote\footnote
\newcommand\nextToken\relax

\renewcommand\footnote[1]{%
    \oldFootnote{#1}\futurelet\nextToken\isFootnote}
    
\newcommand\isFootnote{%
    \ifx\footnote\nextToken\textsuperscript{,}\fi}

\makeatother

\newcommand{\fig}[1]{figure~\ref{#1}}
\newcommand{\tab}[1]{table~\ref{#1}}

\newcommand{\sect}[1]{section~\ref{#1}}

\begin{document}

\title{
    {\Large{}\vspace{-1.0cm}} \hfill {\normalsize{}IFIC/19-40}\\*[10mm]
    {\huge{}Radiative type-I seesaw neutrino masses}{\Large{}\vspace{0.5cm}}
}
  
\date{}

\author[1]{{\large{}Carolina Arbel\'aez}\thanks{E-mail: \href{mailto:carbela.arbelaez@usm.cl}{carbela.arbelaez@usm.cl}}}
\author[1]{{\large{}Antonio E. C\'arcamo}\thanks{E-mail:  \href{mailto:antonio.carcamo@usm.cl}{antonio.carcamo@usm.cl}}}
\author[2]{{\large{}Ricardo Cepedello}\thanks{E-mail:  \href{mailto:ricepe@ific.uv.es}{ricepe@ific.uv.es}}}
\author[2]{{\large{}Martin Hirsch}\thanks{E-mail:  \href{mailto:mahirsch@ific.uv.es}{mahirsch@ific.uv.es}}}
\author[1]{{\large{}Sergey Kovalenko}\thanks{E-mail:  \href{mailto:sergey.kovalenko@usm.cl}{sergey.kovalenko@usm.cl}}}

\affil[1]{\small Universidad T\'ecnica Federico Santa Mar\'ia and
  Centro Cient\'ifico-Tecnol\'ogico de Valpara\'iso \protect\\ Casilla
  110-V, Valpara\'iso, Chile}

\affil[2]{\small AHEP Group, Instituto de F\'isica Corpuscular, CSIC -
  Universitat de Val\`encia \protect\\
  Calle Catedr\'{a}tico Jos\'{e} Beltr\'{a}n, 2 E-46980
  Paterna, Spain }

\maketitle


\begin{abstract}
  We discuss a radiative type-I seesaw. In these models, the radiative
  generation of Dirac neutrino masses allows to explain the smallness
  of the observed neutrino mass scale for rather light right-handed
  neutrino masses in a type-I seesaw. We first present the general
  idea in a model independent way. This allows us to estimate the
  typical scale of right-handed neutrino mass as a function of the
  number of loops. We then present two example models, one at one-loop
  and another one at two-loop, in which we discuss neutrino masses and lepton
  flavour violating constraints in more detail. For the two-loop
  example, right-handed neutrino masses must lie below 100 GeV, thus
  making this class of models testable in heavy neutral lepton searches.
  
\end{abstract}

\newpage{}

\section{Introduction} \label{sec:intro}

\begin{figure}[h!]
    \centering
    \includegraphics[scale=1]{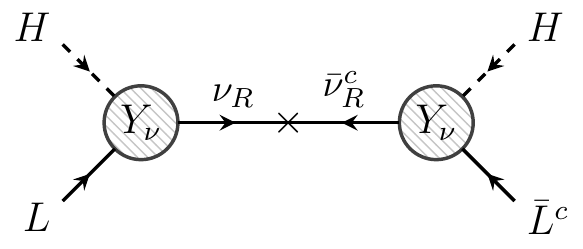}
    \caption{Effective type-I seesaw. The neutrino mass is suppressed
      by the Majorana mass of $\nu_R$ and by the square of the Dirac
      Yukawa term $Y_\nu$ which is generated effectively, see
      Eq.~\eqref{eq:Yuk}.}
    \label{fig:SeesawEff}
\end{figure}

The simplest possibility to generate the Weinberg operator
\cite{Weinberg:1979sa}, 
\begin{eqnarray}\label{eq:Weinberg-1}
&&{\cal O}^W = \frac{1}{\Lambda}LLHH,
\end{eqnarray}
is certainly the type-I seesaw mechanism
\cite{Minkowski:1977sc,Yanagida:1979as,Mohapatra:1979ia}
  given by the diagram in \fig{fig:SeesawEff}. In the classical type-I
  seesaw the Yukawa vertices are point-like $Y_{\nu} \bar{L} H
  \nu_{R}$ and the smallness of the neutrino masses is controlled by a
  large Majorana mass, $\Lambda\sim M_R$, of the right-handed
  neutrinos $\nu_{R}$.

After the electroweak symmetry breaking with the Higgs vacuum
expectation value (vev), $v\equiv\vev{H^0}$, the Weinberg operator
(\ref{eq:Weinberg-1}) leads to the light active neutrino Majorana mass
terms.  In one generation notation, the active neutrino mass is then
given by the well-known relation
\begin{equation} \label{eq:sstI}
    m_\nu \approx m_D^2/M_R, \ \mbox{with}\ m_{D} = Y_{\nu} \langle H^{0}\rangle.
\end{equation}
Assuming that the Yukawas entering $m_D$ take values order ${\cal
  O}(1)$ current neutrino data \cite{deSalas:2017kay} would then point
to $M_R \sim 10^{(14-15)}$ GeV. This setup, apart from being able to
explain neutrino oscillation data, leads only to one experimentally
``testable'' prediction: Neutrinoless double beta decay should be
observed at some level, for reviews on $0\nu\beta\beta$ decay see for
example \cite{Avignone:2007fu,Deppisch:2012nb}.

Here, instead we discuss a simple idea that allows for a
  much lower scale $M_R$, even for all involved Yukawa couplings order
  ${\cal O}(1)$, by generating the Dirac mass term corresponding to
  the Yukawa vertices in \fig{fig:SeesawEff} {\em effectively}.  To
  this end one can claim that the elementary Yukawa coupling is
  forbidden by some symmetry, which being softly broken allows one to
  generate these vertices at certain loop level directly or via higher
  dimension effective operators of the form
\begin{eqnarray}\label{eq:Eff-Operator-1}
&& \frac{\kappa}{M^{2n}} \bar{L} H \nu_{R} \left( H^{\dagger} H \right)^{n},
\end{eqnarray}
where $M$ is the scale of new physics underlying these operators,
supposedly somewhere above the electroweak scale, and $\kappa$ is a
loop suppression factor.  The Dirac mass term is generated by the
operator \eqref{eq:Eff-Operator-1} after the electroweak symmetry
breaking. We assume that only the SM Higgs acquires vev, though it is
straightforward to generalize this to non-SM Higgses with vev's as
well.  Then the resulting effective Yukawa couplings would be
suppressed as
\begin{equation} \label{eq:Yuk}
    Y_\nu \sim \left( \frac{1}{16\pi^2} \right)^\ell  \left( \frac{v^{2}}{M^{2}} \right)^{n},
\end{equation}
where $\ell$ is the number of loops in the diagram generating the
operator \eqref{eq:Eff-Operator-1}.

As $Y_\nu$ is generated effectively, it can be naturally small, while
all couplings arising in the UV complete theory can take values order
${\cal O}(1)$.

As shown in the next section, right-handed neutrino masses of
order of the electroweak scale are easily possible in this setup. Such
moderately heavy right-handed neutrinos could be searched for in
accelerator based experiments via displaced vertices.  The topic of
"long-lived light particles" (LLLPs) has attracted much attention in
the recent literature \cite{Alimena:2019zri}.  A number of recent
experimental proposals
\cite{Curtin:2018mvb,Gligorov:2017nwh,Feng:2017uoz,Gligorov:2018vkc}
could search for this signal.  Sensitivity estimates for right-handed
neutrinos for these experiments can be found in
\cite{Helo:2018qej,Dercks:2018wum}, for the LHC main experiments in
\cite{Cottin:2018kmq,Liu:2019ayx,Drewes:2019fou}.

As we already mentioned, in order to forbid tree-level Dirac Yukawa
couplings, it is necessary to postulate some additional symmetry
beyond the standard model (SM) gauge group. This symmetry could be
either gauged or discrete. For simplicity, in our model constructions
we will concentrate on discrete symmetries. Starting with a $Z_4$
symmetry, which gets softly broken to an exact remnant $Z_2$
symmetry. Thus, the same symmetry responsible for explaining the
smallness of the neutrino mass is able to stabilize a dark matter
candidate too.

In our setup neutrinos are Majorana particles. However, our
constructions have some overlap with recent papers on Dirac neutrinos.
Some general considerations on how to obtain small Dirac neutrino
masses have been discussed in \cite{Ma:2016mwh}.  Systematic studies
of one-loop (and two-loop) Dirac neutrino masses were given in
\cite{Yao:2017vtm}, (\cite{CentellesChulia:2019xky}). Also the
generation of $d=6$ Dirac neutrino masses has been considered
\cite{CentellesChulia:2018bkz}.

The rest of this paper is organized as follows. In the next section,
we will discuss the radiative generation of neutrino Dirac couplings
from a model-independent point of view. This allows us to estimate the
typical scales for the Majorana mass of neutrinos as a function of
the loop level, at which the Dirac couplings are generated.  In
section \ref{sec:examples} we will discuss then two concrete example
models, at one-loop and two-loop level. For these we will
estimate in more detail the neutrino masses, discuss possible
constraints from lepton flavor violation and then turn briefly to
dark matter.  We will then close with a short summary and outlook.

\section{Radiative type-I seesaw} \label{sec:radiative}

In this brief section we will discuss the radiative generation of
neutrino Dirac Yukawa couplings from a model-independent point of
view.  Here we consider only the $d=4$ Dirac mass operator $L H \bar
\nu_R$ generated via loops. The mass of the light active neutrinos
arises then from the diagram depicted in \fig{fig:SeesawEff} and
is given by eq. \eqref{eq:sstI}.

For simplicity, we will limit ourselves to discussing the
phenomenologically unrealistic case of one massive neutrino with no
hierarchy or flavour structure for the Yukawas. This is sufficient for
discussing the parameter dependence, extending to three generations of
active neutrinos is straightforward. The Dirac Yukawa, $Y_\nu$, can be
parametrized in general in terms of five exponents
$(\ell,\alpha,\beta,\gamma,\delta) \in \mathbb{N}^+$, whose values will
depend on the specific UV complete realization of the operator $Y_\nu
\, L H \bar \nu_R$, as:
\begin{equation} \label{eq:paramYuk}
    Y_\nu \sim \left( \frac{1}{16 \pi^2} \right)^\ell \left( \frac{m_\tau}{v} \right)^\alpha \left( \frac{M_F}{\Lambda} \right)^\beta \left( \frac{\mu}{\Lambda} \right)^\gamma \epsilon^\delta.
\end{equation}
This corresponds to generating effectively the Yukawa via a diagram
with
\begin{itemize}
    \item $\ell$ loops;
    \item $\alpha$ insertions of SM Yukawas. Unless the UV model 
      allows for a top-quark in the loop, 
      this  corresponds to a suppression of typically $\sim 10^{-2}$,
      from $Y_{\tau}^{SM}$ (or $Y_{b}^{SM}$);
    \item $\beta$ mass insertions of new (vector-like) fermions,
      not part of the SM, all set to $M_F$ for simplicity;
    \item $\gamma$ dimensionful couplings in the scalar sector, i.e.
      trilinear scalar couplings;
    \item $\delta$ dimensionless couplings, for instance Yukawas or
      four-point scalar couplings.
\end{itemize}
Not all possible sets of exponents can be realized in a UV complete
model which is genuine, i.e. give the dominant contribution to the
neutrino mass. For example, for the most simple case of an one-loop
Dirac mass term, there are only two genuine diagrams
\cite{Yao:2017vtm} with one or two mass insertions and, at least,
three couplings. So, for $\ell=1$ it is not possible to generate a
genuine diagram with, for instance, $\alpha,\beta > 2$. The possible
combinations of $(\alpha,\beta,\gamma,\delta)$ can be deduced from the
systematic studies of radiative Dirac models given in
\cite{Yao:2017vtm,CentellesChulia:2019xky}.

For our numerical estimates, we will assume that all couplings are in
the perturbative regime, i.e. $\epsilon \lesssim 1$.\footnote{It is
  often argued that perturbativity only requires Yukawa couplings to
  be $Y \lesssim \sqrt{4\pi}$. However, saturating this limit would
  imply that higher order contributions are (at least) equally
  important than the leading order (that we consider), thus rendering
  estimates effectively inconsistent.}  If $\mu$ is a trilinear
coupling between some BSM scalar and the Higgs, it enters in the
calculation of the stability of the Higgs potential, i.e. it will
induce a modification of the quartic Higgs coupling at one-loop
level. We will thus also assume that $\mu \lesssim m_S \equiv \epsilon
\, m_S$, in order not to run into problems with the SM Higgs sector.
With these considerations the light neutrino mass can be written in
terms of the same five exponents, using the seesaw relation \eqref{eq:sstI},
\begin{equation} \label{eq:param}
    m_\nu \sim \left( \frac{1}{16 \pi^2} \right)^{2\ell} \frac{v^2}{M_R} \left( \frac{m_\tau}{v} \right)^{2\alpha} \left( \frac{M_F}{\Lambda} \right)^{2\beta} \left( \frac{m_S}{\Lambda} \right)^{2\gamma} \epsilon^{2(\gamma+\delta)}.
\end{equation}
As this equation shows, neutrino masses generated from this class of
models will be very suppressed. If, for instance, the Dirac neutrino
mass arises at two-loop order, then $m_\nu$ will effectively come from
a four-loop diagram with an extra suppression due to the Majorana
scale $M_R$. Thus, for relatively low masses of the order of TeV
and couplings order one, a reasonable neutrino mass can be obtained
easily.

A rough, but conservative limit on the Majorana mass scale, can be obtained
setting all masses in the loop to the same scale $\Lambda = M_F = m_S$. 
Conservatively taking $\epsilon = 1$, we find
\begin{equation} \label{eq:ParMnu2}
    m_\nu \sim \left( \frac{1}{16 \pi^2} \right)^{2\ell} \frac{v^2}{M_R} \left( \frac{m_\tau}{v} \right)^{2\alpha}.
\end{equation}
Note, that the scale $\Lambda$ does not appear in this simple case in the
expression for $m_{\nu}$. This is to be expected, given the $d=4$
nature of the neutrino Dirac coupling. Taking as reference scale the
atmospheric neutrino mass $\sqrt{|\Delta m_{31}^2|} \approx 0.05$ eV,
we can set upper limits on $M_R$ as function of the exponents $\ell$
and $\alpha$. Limits are given in \tab{tab:GenLim} up to three-loops
and two SM Yukawa insertions.  The numbers given correspond to
couplings order one.

\begin{table}[h!]
    \begin{center}
        \begin{tabular}{ c|*{3}{|c} }
            \xrowht[()]{10pt}
            $M_R$ & $\alpha=0$ & $\alpha=1$ & $\alpha=2$ \\
            \hline\hline
            \xrowht[()]{14pt}
            $\ell=1$ & $2\times 10^{10}$ GeV & $2\times 10^{6}$ GeV & $2\times 10^2$ GeV \\
            \hline
            \xrowht[()]{14pt}
            $\ell=2$ & $10^{6}$ GeV & $10^2$ GeV & $9\times 10^{-3}$ GeV \\
            \hline
            \xrowht[()]{14pt}
            $\ell=3$ & $4\times 10^1$ GeV & $4\times 10^{-3}$ GeV & $4\times 10^{-7}$ GeV
        \end{tabular}
    \end{center}
    \caption{Estimated values for $M_R$ needed to fit a neutrino mass
      of $0.05$ eV with couplings order one for different realizations
      of the Dirac mass operator $L H \bar \nu_R$, considering $\ell$
      loops and $\alpha$ SM Yukawa insertions. These mass scales
      constitute a rough, but conservative upper limit for $M_R$ for
      each class of models parametrized by the exponents $\ell$ and
      $\alpha$ in \eqref{eq:ParMnu2}.}
    \label{tab:GenLim}
\end{table}

Obviously, $M_R$ decreases very fast as $\alpha$ or $\ell$
increase. This is due to the fact that for Majorana neutrinos $m_\nu$
depends quadratically on $Y_\nu$, rather than linearly. For
$\alpha=1$ and $l=2$ one finds a scale of $M_R \sim 10^2$ GeV
and similar numbers for $\alpha=2$ and $l=1$ or $\alpha=0$ and $l=3$.
These are the phenomenologically most interesting cases. 

Apart from the {\em upper} limit on the Majorana mass coming from the
neutrino mass scale, lower limits on $M_R$ can be set from big bang
nucleosynthesis \cite{Deppisch:2015qwa} and the effective number of
neutrinos in the early universe $\Delta N_{eff}$
\cite{Gariazzo:2019gyi}. These limits depend on the mixing angle
between the right-handed and the active neutrinos (as a function of
the mass $M_R$). For our class of models, as for the ordinary type-I
seesaw, one expects $M_R \gtrsim (0.1-1)$ GeV, from these considerations
\cite{Deppisch:2015qwa,Gariazzo:2019gyi}. This constrains 
significantly the space of possible models to only those with 
three loops or less and at most two SM mass insertions (for 
the case of 1-loop). In the next section, we will therefore discuss
two model examples in more detail: a one-loop and a two-loop
model.

\section{Examples of models} \label{sec:examples}

In this section we show two simple models where the Dirac mass is
generated at one- and two-loops, both containing a stable dark matter
candidate, which participates in the loop. We give an estimate of the
neutrino mass scale involved for a simplified benchmark, as well as an
insight to the phenomenological constraints coming from charged lepton
flavour violating processes.

\subsection{One-loop Dirac mass} \label{sec:oneloop}

The particle spectrum of the model and their assignments under the SM
gauge and the $Z_4$ discrete symmetry are shown in
\tab{tab:fields}. Notice that we have assumed a $Z_4$
symmetry, which is softly broken down to the preserved $Z_2$ symmetry,
in order to guarantee that the Dirac neutrino mass matrix is generated
at one-loop level. The scalar sector of the model is composed of the
SM Higgs doublet $H$, the inert $SU(2)_L$ scalar doublet $\eta$ and
the electrically charged gauge singlet scalar $S^-$. In addition, the
SM fermion sector is extended by the inclusion of a right-handed
Majorana neutrino $\nu_{R}$ \footnote{We repeat,
  that we are interested here only in a rough estimate for the
  neutrino mass scale. For phenomenological reasons, one would need
  indeed at least two right-handed neutrinos that generate the solar and
  atmospheric neutrino mass. Since fits of the seesaw type-I to
  neutrino data are straightforward and have been done many times in
  the literature, we do not repeat these details here.} and the vector
like charged leptons $\chi_L$ and $\chi_R$. The relevant terms
for the neutrino mass take the form,
\begin{eqnarray} \label{eq:Lag-Y}
    - \mathcal{L}_{Y} &=& Y_e \, L H^{\dagger} e^c + Y_{L} \, L \eta^{\dagger} \chi_{L}
    + Y_{R} \, \overline{\chi_{R}} S^{+} \overline{\nu_{R}} + \text{h.c.},
\\ \label{eq:Lag-M}
    \mathcal{L}_M &=& M_{R} \, \overline{\nu _{R}^{c}} \nu_{R}
    + M_{\chi} \, \overline{\chi_{R}} \chi _{L} + \text{h.c.},
\end{eqnarray}
flavour indices and $SU(2)$ contractions have been suppressed for
brevity.

\begin{figure}
    \centering
    \begin{tikzpicture}[line width=1. pt, scale=1.5]
        \draw[fermion] (-1.5,0) -- (-1,0)
            node[midway, below] {$L_\alpha$};
        \draw[fermionbar] (-1,0) -- (0,0)
            node[midway, below] {$\chi_L$}
            node[] {$\times$};
        \draw[fermion] (0,0) -- (1,0)
            node[midway, below] {$\bar \chi_R$};
        \draw[fermion] (1,0) -- (1.5,0)
            node[midway, below] {$\nu_{R\beta}$};
        \draw[scalar] (-1,0) arc (180:93:1)
            node[midway,left] {$\eta$};
        \draw[scalar] (1,0) arc (0:87:1)
            node[midway,right] {$S^-$};
        \draw[scalarbar] (0,1.1) -- (0,1.5)
            node[above,left] {$H$};
        \node at (0,1) {$\mathbf{\otimes}$};
    \end{tikzpicture}
    \caption{One-loop Dirac neutrino mass. The diagram is
      realized when the $Z_4$ is softly broken (denoted by the symbol
      $\otimes$). As the symmetry is broken in two units, the diagram
      is still invariant under a remnant $Z_2$ of $Z_4$.}
    \label{fig:model 1}
\end{figure}
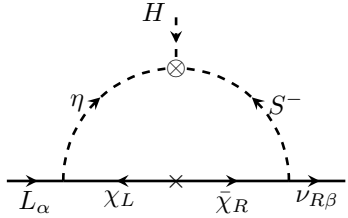

The terms above generate the Dirac neutrino mass matrix at one-loop
level through the diagram shown in \fig{fig:model 1} provided the
following $Z_4$ trilinear soft breaking term is added to the scalar
potential,
\begin{equation}
\mathcal{V} \supset \mu_S \, H \eta S^{-} + \text{h.c.}
\end{equation}
The softly broken $Z_{4}$ guarantees that the Dirac mass term is
forbidden at tree-level but generated by loops, i.e. that the diagram
is genuine (non-reducible) \cite{Yao:2017vtm}.

\begin{table}
\begin{center}
\begin{tabular}{| c || c | c || c |}
  \hline
  \hspace{0.1cm} Fields \hspace{0.1cm}  &  $SU(3)_C \times SU(2)_L \times U(1)_Y$  &  \hspace{0.2cm} $Z_4$ \hspace{0.2cm}  &  Residual $Z_2$  \\
\hline \hline
                   $L$ & ($\mathbf{1}$, $\mathbf{2}$, -1/2) &  $1$ &  $1$ \\
                 $e^c$ & ($\mathbf{1}$, $\mathbf{1}$,    1) &  $1$ &  $1$ \\
               $\nu_R$ & ($\mathbf{1}$, $\mathbf{1}$,    0) & $-1$ &  $1$ \\
                   $H$ & ($\mathbf{1}$, $\mathbf{2}$,  1/2) &  $1$ &  $1$ \\
\hline
  ($\chi_{L}$, $\chi_{R}$) & ($\mathbf{1}$, $\mathbf{1}$,    1) &  ($i$, $i$) &  ($-1$, $-1$) \\
                    $\eta$ & ($\mathbf{1}$, $\mathbf{2}$,  1/2) &         $i$ &          $-1$ \\
                     $S^-$ & ($\mathbf{1}$, $\mathbf{1}$,   -1) &         $i$ &          $-1$ \\
    \hline
  \end{tabular}
\end{center}
\caption{Particle content of the example model that generates the
  one-loop diagram of \fig{fig:model 1} once the $Z_4$ is softly
  broken by the trilinear term $H \eta S^-$. After the breaking of
  $Z_4$ a remnant $Z_2$ is exactly conserved.}
 \label{tab:fields}
\end{table}%

The Dirac mass term can be computed directly from the diagram in 
\fig{fig:model 1} given the Lagrangians eqs \eqref{eq:Lag-Y} and 
\eqref{eq:Lag-M} and the soft-breaking term. In the mass insertion
approximation and, for simplicity, setting all the masses of the internal scalars,
as well as the soft-breaking parameter $\mu_S$, to $m_S$, one finds:
\begin{equation} 
    m_D \approx \frac{1}{16\pi^2} \frac{v m_S}{M_\chi} Y_L Y_R \; \mathcal{I}_1(m_S^2/M_\chi^2).
\end{equation}
The loop integral $\mathcal{I}_1(x)$ can be written in terms of the Passarino-Veltman $B_0$ function \cite{Passarino:1978jh} as,
\begin{equation} 
    \mathcal{I}_1(x) = \frac{1}{1-x} \left[ B_0(0,1,x) - B_0(0,x,x) \right].
\end{equation}

The mass scale of the lightest active neutrino can be directly estimated
through the seesaw approximation as,
\begin{equation} \label{eq:mnu 1}
m_{\nu} \sim \left(\frac{1}{16\pi^2} \right)^2 Y_{L}^2Y_{R}^2 \, \frac{v^2 m^2_{S}}{M^2_{\chi}M_{R}} \, [\mathcal{I}_1(m_{S}^2/M_{\chi}^2)]^2
\end{equation}

\begin{figure}
\begin{center}
\includegraphics[width=0.7\textwidth]{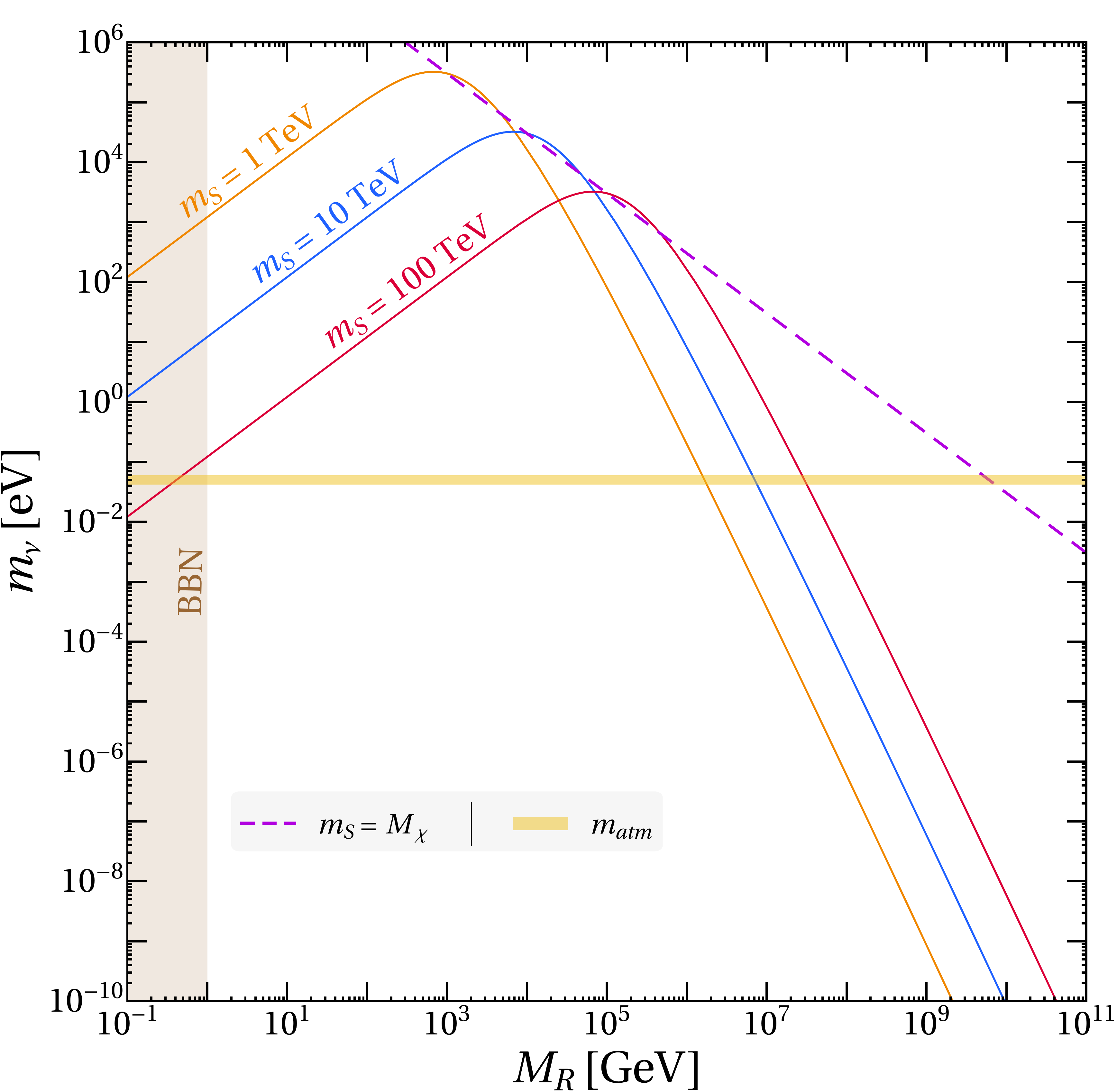} 
\caption{One-loop neutrino mass scale. The dashed line corresponds to
  the case where $m_{S}=M_{\chi}$, while the solid lines depicted the
  case where $M_{\chi}=M_{R}$ for different scalar masses. The Yukawas
  $Y_{L}$ and $Y_{R}$ are set to 1. Big bang nucleosynthesis (BBN)
  excludes $M_R >(0.1-1)$ GeV, depending on mixing, for these class of
  models \cite{Deppisch:2015qwa}.}
\label{fig:oneloopmnu}
\end{center}
\end{figure}

This mass scale as a function of $M_{R}$ is plotted in
\fig{fig:oneloopmnu}. Two different benchmarks with $M_{\chi}=M_{R}$
and $m_{S} = M_{\chi}$ are represented by the solid and dashed lines
respectively. For both cases, we can observe that the neutrino mass is
strongly suppressed even for small values on $M_{R}$. In the $m_{S} =
M_{\chi}$ scenario, the neutrino mass falls as $\sim 1/M_{R}$
independently of the one-loop internal scalar masses. Moreover, in the
$M_{\chi}=M_{R}$ scenario, the neutrino mass is a function of both
mass scales $m_{S}$ and $M_{R}$. It behaves as $M_R$ or $1/M_R^3$
depending on which of these two scales dominate the loop.

The window of allowed $M_{R}$ values which could fit the neutrino
oscillation scale $m_{\nu} \sim 0.05$ eV becomes narrower for larger
masses $m_{S}$. Note that in \fig{fig:oneloopmnu} the neutrino mass is
plotted for order one couplings. Consequently, the points with
a neutrino mass lying roughly below the atmospheric scale is
phenomenologically non-viable, as it would require couplings larger
than one (non-perturbative) to give a reasonable mass scale.

Current upper limits on lepton flavour violating (LFV) decays such as $\mu
\rightarrow e\gamma$ can provide constraints on the parameters of our
model. These depend on specific choices for the Yukawas $Y_L$ and
$Y_R$. As eq. \eqref{eq:mnu 1} shows, $m_{\nu}$ depends on the product
of these couplings, while LFV decays are mostly sensitive to $Y_L$ only.
There are then two extreme cases: (i) Choose $Y_L \simeq 1$  and
fit $Y_R$ to $m_{\nu}$ as function of the other model parameters and
(ii)  $Y_R \simeq 1$  and fit $Y_L$. Case (i) is very similar to the 
situation in our two-loop model (see \sect{sec:twoloop}), and thus we will discuss
the details in the next section. For case (ii) on the other
hand, we found that LFV limits do not impose interesting limits on
our one-loop model.

The residual $Z_2$ symmetry ensures that the lightest of the fields
running inside the loop will be stable. In order to not run into
conflict with cosmology and to provide a good dark matter candidate,
one should force the neutral component of the doublet $\eta$ to be the
lightest of the loop particles. Similar DM candidates have been
studied in the literature\footnote{See for instance the well-known
  \textit{Inert Doublet Model} \cite{Deshpande:1977rw} or the
  \textit{Scotogenic model} \cite{Ma:1998dn}.}. Considering $\eta$ as
the only source of dark matter, the observed relic density, together
with direct detection limits and the constraints on the invisible
width of the Higgs boson severely limit its mass to lie either around
$m_{h}/2 \simeq 62.5$ GeV, in a small region around $m_\eta\simeq 72$
GeV or above $m_\eta \gtrsim 500$ GeV \cite{Eiteneuer:2017hoh}.

\subsection{Two-loop Dirac mass} \label{sec:twoloop}

Analogously to the first example, we build a two-loop radiative seesaw
model breaking softly a $Z_4$ discrete group to an exact $Z_2$
symmetry. The particle content and their transformation properties
under the SM gauge and the $Z_4$ discrete symmetry are shown in
\tab{tab:fields2loops}. We again include a right-handed Majorana
neutrino $\nu_{R}$.

The relevant terms of the Lagrangian and the scalar sector invariant
under $Z_4$ are,
\begin{eqnarray}
    -\mathcal{L}_{Y} &=& Y_e \, L H^{\dagger} e^{c} + Y_L \, \overline{F_L} \eta_{2} \overline{e^c} + Y_R \, \overline{\nu_{R}} \eta_{2} F_{L} + \text{h.c.},
    \\
    \mathcal{L}_{M} &=& M_{R} \, \overline{\nu_{R}^{c}} \nu_{R} + M_{F} \, \overline{F_R} F_{L} + \text{h.c.},
    \\
    \mathcal{V} &\supset& \lambda \, \eta_{1}^{\dagger}H \eta_{2}^{\dagger} H + \text{h.c.},
\end{eqnarray}

\begin{table}
\begin{center}
\begin{tabular}{| c || c | c || c |}
  \hline
  \hspace{0.1cm} Fields \hspace{0.1cm}  &  $SU(3)_C \times SU(2)_L \times U(1)_Y$  &  \hspace{0.2cm} $Z_4$ \hspace{0.2cm}  &  Residual $Z_2$  \\
\hline \hline
                   $L$ &  ($\mathbf{1}$, $\mathbf{2}$, -1/2) &   $1$ &  $1$ \\
                 $e^c$ &  ($\mathbf{1}$, $\mathbf{1}$,    1) &   $1$ &  $1$ \\
               $\nu_R$ &  ($\mathbf{1}$, $\mathbf{1}$,    0) &  $-1$ &  $1$ \\
                   $H$ &  ($\mathbf{1}$, $\mathbf{2}$,  1/2) &   $1$ &  $1$ \\
\hline
    ($F_{L}$, $F_{R}$) &  ($\mathbf{1}$, $\mathbf{2}$, -1/2) &  ($i$, $i$) &  ($-1$, $-1$) \\
              $\eta_1$ &  ($\mathbf{1}$, $\mathbf{2}$,  1/2) &        $-i$ &          $-1$ \\
              $\eta_2$ &  ($\mathbf{1}$, $\mathbf{2}$,  1/2) &         $i$ &          $-1$ \\
    \hline
  \end{tabular}
\end{center}
\caption{Particle content of the example model that generates the two-loop
  diagram of \fig{fig:model 2} once the $Z_4$ is softly broken by
  the term $\eta_2^\dagger \eta_1$. After the breaking of $Z_4$ a
  remnant $Z_2$ is conserved.}
 \label{tab:fields2loops}
\end{table}%

\begin{figure}
    \centering
    \begin{tikzpicture}[line width=1. pt, scale=1.5]
        \draw[fermion] (-1.5,0) -- (-1,0)
            node[midway, below] {$L_\alpha$};
        \draw[fermionbar] (-1,0) -- (0,0)
            node[midway, below] {$e^c$};
        \draw[fermion] (0,0) -- (1,0)
            node[midway, below] {$F_L$};
        \draw[fermion] (1,0) -- (1.5,0)
            node[midway, below] {$\nu_{R\beta}$};
        \draw[scalar] (-1,0) arc (180:90:1)
            node[midway,left] {$H$};
        \draw[scalar] (0,1) arc (90:45:1)
            node[midway,above] {$\eta_1$};
        \draw[scalarbar] (1,0) arc (0:40:1)
            node[midway,right] {$\eta_2$}
            node[rotate=45] {$\otimes$};
        \draw[scalarbar] (0,0) -- (0,1)
            node[midway,left] {$\eta_2$};
        \draw[scalarbar] (0,1) -- (0,1.5)
            node[above,left] {$H$};    
    \end{tikzpicture}
    \caption{Two-loop Dirac neutrino mass. The diagram is
      realized when the $Z_4$ is softly broken (denoted by the symbol
      $\otimes$). As the symmetry is broken in two units, the diagram
      is still invariant under a remnant $Z_2$.}
    \label{fig:model 2}
\end{figure}
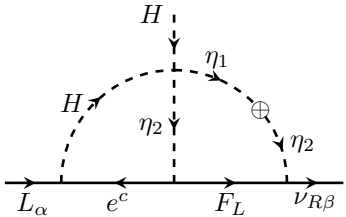

An effective Dirac term is generated once the $Z_4$ symmetry is softly
broken in the scalar sector by the term,
\begin{equation}
-\mathcal{L}_{\text{soft}} = \mu_{12}^{2} \, \eta_{2}^{\dagger} \eta_{1} + \text{h.c.}
\end{equation}
A Dirac mass appears at the two-loop level, as depicted in
\fig{fig:model 2}, which can be expressed in the mass insertion
approximation, assuming no flavour structure in the Yukawa couplings,
as
\begin{equation}
m_{D} \approx \left( \frac{1}{16 \pi^2} \right)^2 \lambda \, Y_{e} Y_{L} Y_{R} \frac{v \, \mu_{12}^2 }{ M^2_F } \, \mathcal{I}_2 ( m_{S}^{2} / M_{F}^{2} )
\end{equation}
with the $\mathcal{I}_2(x)$ a dimensionless two-loop
function. $\mu_{12}$ is the soft breaking mass term depicted by
$\otimes$ in \fig{fig:model 2}. For simplicity, we set all the masses
of the new internal scalars to $m_{S}$. Taking into account that the
main contribution of the SM Yukawa $Y_{e}$ would be $m_{\tau}/v$, the
mass scale of the lightest active neutrino is directly estimated
through the seesaw approximation as,
\begin{equation} \label{eq:mnu 2}
m_{\nu} \sim \left( \frac{1}{16 \pi^2} \right)^4 \lambda^2 Y_L^2 Y_R^2 \, \frac{ m_{\tau}^2 m_{S}^4 }{ M_F^4 M_{R} } \, \left[ \mathcal{I}_2(m_{S}^2/M_{F}^2) \right]^2,
\end{equation}
where as before, we have set $\mu_{12} = m_S$. $\mathcal{I}_2$ can be
written in terms of simple two-loop integrals for which analytical
solutions are known \cite{Martin:2016bgz}. We do not give here its
decomposition for brevity, though it can be found in the literature
\cite{Sierra:2014rxa}.

\begin{figure}
\begin{center}
\begin{tabular}{cc}
\hspace{-1cm}
\includegraphics[width=0.7\textwidth]{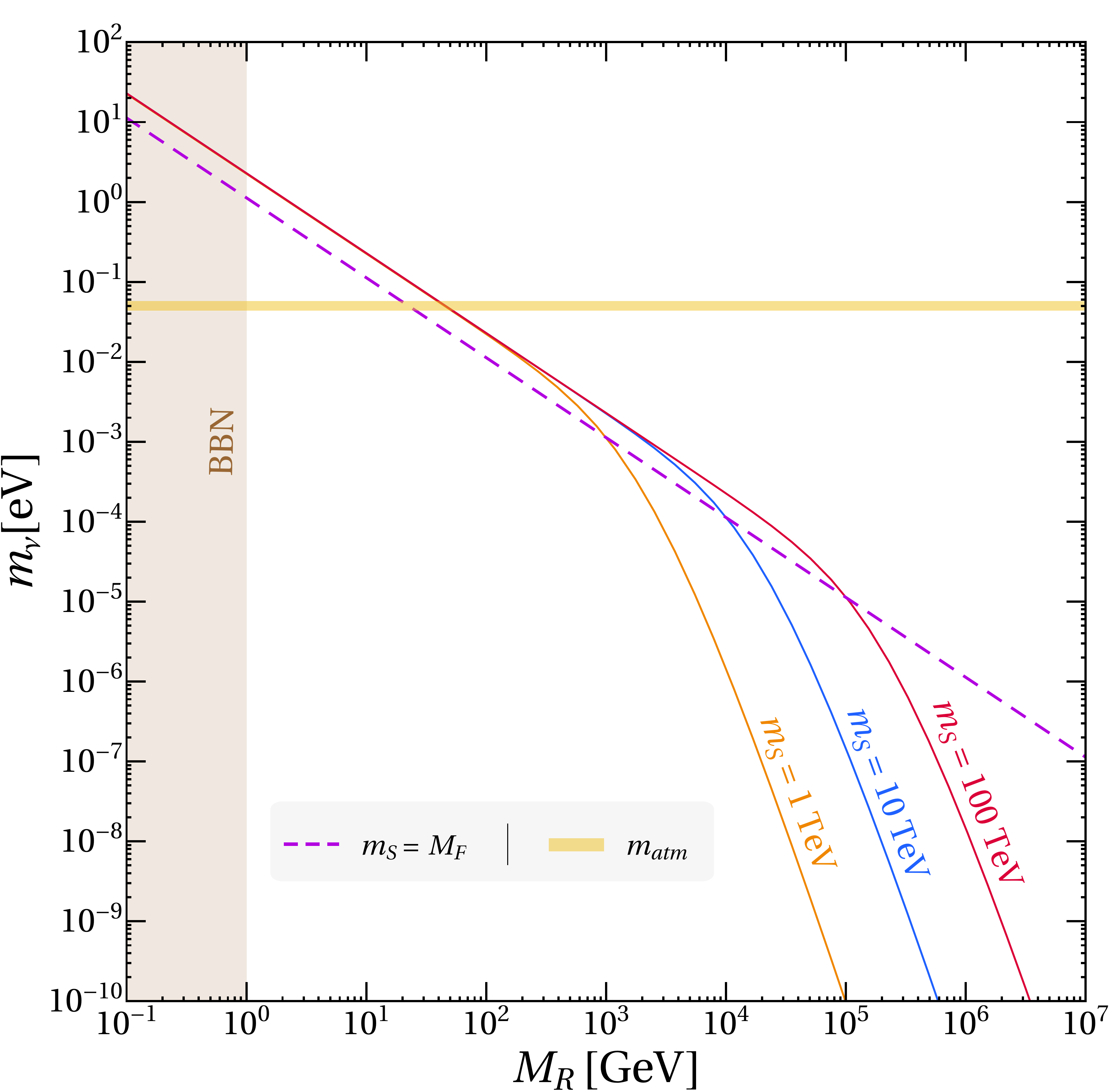}
\end{tabular}
\caption{Two-loop neutrino mass scale assuming that $m_{S}=M_{F}$
  and $M_{F}=M_{R}$, depicted in dashed and solid lines
  respectively. All dimensionless couplings are set to 1 and the BBN
  exclusion region is indicated in the left.}
\label{fig:mass plot 2}
\end{center}
\end{figure}

The neutrino mass scale, eq. \eqref{eq:mnu 2}, as a function of
$M_{R}$ is plotted in \fig{fig:mass plot 2}. We consider two different
approximations: $M_{F}=m_{S}$ and $M_{F}=M_{R}$, represented by the
dashed and solid lines respectively. As expected from
\tab{tab:GenLim}, the neutrino mass is more strongly suppressed
compared with the one-loop model described previously. For the case
$m_{S}=M_{F}$ the Dirac Yukawa is independent of the scale,
consequently the neutrino mass falls simply as $\sim 1/M_{R}$. On the
other hand, in the scenario where $M_{F}=M_{R}$, this same behaviour
is reproduced when $m_{S}$ dominates, while for values of $M_{R} >
m_S$, the neutrino mass follows the curve $1/M_R^5$.

Given the suppression factor $(m_\tau/v)^2 \sim 10^{-4}$, and if we
take into account the limit coming from cosmology (BBN), the range of
allowed values of $M_{R}$ which can fit the neutrino oscillation scale
$m_{atm} \sim 0.05$ eV is considerably limited. For $m_S>10^2$ GeV,
$M_{R}$ has to be $M_{R} \lesssim 10^{2}$ GeV. This makes the model
testable in future heavy neutral lepton searches.

We mention again that the remnant $Z_2$ symmetry stabilises the
lightest of the fields odd under this symmetry. Fermionic dark matter
coming from a doublet is ruled out by direct detection experiments
\cite{Cirelli:2005uq}, while for the scalar inert doublet the same
limits described in the previous section apply.

\begin{figure}
\begin{center}
\includegraphics[width=0.7\textwidth]{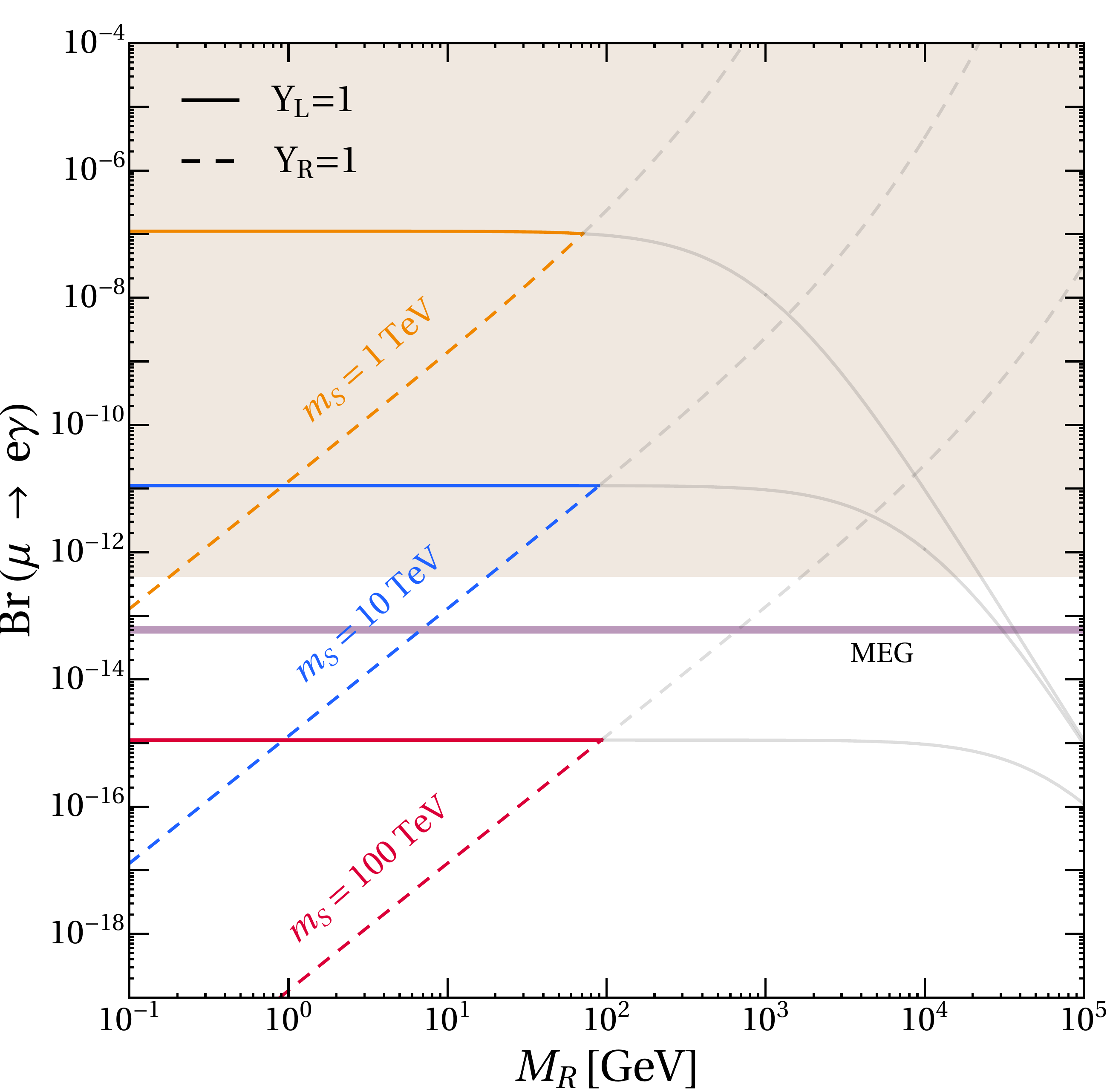}
\caption{Estimate of the branching ratio of $\mu \rightarrow e \gamma$
  as a function of $M_{R}$ for different values of $m_{S}$ fitting the
  neutrino mass to $m_{atm}$. The areas between the coloured lines are
  allowed in this model, see text.  The grey lines represent the
  values of $M_{R}$ where one of the Yukawa couplings becomes 
  non-perturbative in order to fit neutrino oscillation data. The
  shadowed region represents the experimentally excluded area for
  $Br(\mu \rightarrow e \gamma) > 4.2 \times 10^{-13}$
  \cite{TheMEG:2016wtm}, while the purple line corresponds to the
  future prospect limit from MEG collaboration
  \cite{Baldini:2013ke}.}
   \label{fig:dwSST3}
\end{center}
\end{figure}

Turning to LFV processes, \fig{fig:dwSST3} shows the $Br(\mu
\rightarrow e \gamma)$ as a function of $M_{R}$ for two different
scenarios, already mentioned in section \ref{sec:oneloop}: (i) $Y_L
\simeq 1$ and fit $Y_R$ to $m_{\nu}$ or (ii) $Y_R \simeq 1$ and fit
$Y_L$. All other possibilities to choose Yukawas lie between these
extremes. The dominant (one-loop) contribution to $Br(\mu \rightarrow
e \gamma)$ comes always from $Y_L$, which directly connects the new
particles with the SM leptons. For $M_F=M_R$ and $Y_R=1$ the branching
is dominated by the fit of the neutrino mass, eq. \eqref{eq:mnu 2}. The
branching increases as function of $M_R$, as $Y_L$ gets larger
counteracting the suppression of $1/M_R^5$ in the neutrino mass. We
stop the calculation when $Y_L$ grows larger than 1. In contrast, for
$Y_L=1$ there is no dependence from the neutrino mass fit, but a
suppression of $1/M_R^4$ when this mass scale dominates over $m_S$ in
the $\mu \rightarrow e \gamma$ loop function \cite{Lavoura:2003xp}. 
The regions in between those extremes are the regions allowed for
this neutrino mass model.

\section{Conclusions} \label{sec:conclusions}

We have constructed a new realization of the type-I seesaw mechanism
based on radiatively generated Dirac neutrino masses. We showed that
this class of models can naturally generate a small neutrino mass for
 order one couplings and relatively low mass scales. Compared to the
standard type-I seesaw mechanism, for which the Majorana mass scale
should be of the order of the GUT scale, we found viable models even
for $M_R$ below 100 GeV. Parametrizing the neutrino mass in terms of
five integers, we derived for each set of models a conservative limit
on $M_R$ requiring only that they should fit the atmospheric neutrino
mass scale. The strong suppression of the light neutrino
mass with the number of loops, i.e. $(1/16\pi^2)^{2\ell}$, along with
the seesaw Majorana mass suppression, allows remarkably low $M_{R}$
values. This fact makes models with large number of loops (or SM mass
insertions) run into conflict with big bang nucleosynthesis and $\Delta
N_{eff}$, which therefore significantly constrains the space of
possible models.

To illustrate in further detail this idea, we presented two example
models where the Dirac neutrino mass matrix is generated at one- and
two-loop level. The latter lies at the edge of the excluded models. An
extra $Z_4$ symmetry is incorporated to forbid a tree-level Dirac
mass, but broken softly in order to generate the Dirac Yukawa
radiatively. A remnant exact $Z_2$ symmetry is kept stabilising the
lightest of the $Z_2$ charged fields and providing a good dark matter
candidate.

\bigskip

\centerline{\bf Acknowledgements}

\medskip
C.A., A.E.C. and S.K. are supported by Fondecyt (Chile) grants No.~11180722, No.~1170803, No.~1190845 and CONICYT PIA/BASAL FB0821. R.C. and M.H. acknowledge funding
by Spanish grants FPA2017-90566-REDC (Red Consolider MultiDark),
FPA2017-85216-P and SEV-2014-0398 (AEI/FEDER, UE), PROMETEO/2018/165
(Generalitat Valenciana) and FPU15/03158. R.C. is also supported by
Beca Santander Iberoam\'erica 2018/19 and would like to thank the USM
Department of Physics and CCTVal for their hospitality.

\bigskip
 
\appendix

\bibliographystyle{BibFiles/t1}
\bibliography{BibFiles/bibliography}

\end{document}